\documentclass[pdflatex,sn-mathphys]{sn-jnl}

\jyear{2021}%

\theoremstyle{thmstyleone}%
\theoremstyle{thmstyletwo}%

\theoremstyle{thmstylethree}%

\newcommand\aap{A\&A}                
\newcommand\aj{AJ}                   
\newcommand\apj{ApJ}                 

\newcommand{\Oiii}[1]{[\ion{O}{III}] \ensuremath{#1}}

\newcommand{\Oiirec}[1]{\ion{O}{II} \ensuremath{#1}}
\newcommand{\Oii}[1]{[\ion{O}{II}] \ensuremath{#1}}

\newcommand{\Hii}{\ion{H}{II}}

\newcommand{\um}{\ensuremath{\mu\mathrm{m}}}
\newcommand\ion[2]{\textnormal{#1\,\textsc{\lowercase{#2}}}}
\newcommand{\chb}{\ensuremath{c(\mathrm{H}\beta)}}

\newcommand{\kms}{\ensuremath{\mathrm{km}/\mathrm{s}}}

\begin{document}

\title[Reply to M-D]{Reply to: Temperature inhomogeneities in Mrk 71 can not be discarded}


\author*[1]{\fnm{Yuguang} \sur{Chen}}\email{yugchen@ucdavis.edu}

\author[1]{\fnm{Tucker} \sur{Jones}}

\author[1,2]{\fnm{Ryan} \sur{Sanders}}

\author[3,4]{\fnm{Dario} \sur{Fadda}}

\author[3,5]{\fnm{Jessica} \sur{Sutter}}

\author[3,6]{\fnm{Robert} \sur{Minchin}}

\author[1]{\fnm{Erin} \sur{Huntzinger}}

\author[7]{\fnm{Peter} \sur{Senchyna}}

\author[8]{\fnm{Daniel} \sur{Stark}}

\author[9]{\fnm{Justin} \sur{Spilker}}

\author[8]{\fnm{Benjamin} \sur{Weiner}}

\author[10]{\fnm{Guido} \sur{Roberts-Borsani}}

\affil*[1]{\orgdiv{Department of Physics \& Astronomy}, \orgname{University of California, Davis}, \orgaddress{\street{1 Sheilds Avenue}, \city{Davis}, \postcode{95616}, \state{CA}, \country{USA}}}

\affil[2]{\orgdiv{Department of Physics and Astronomy}, \orgname{University of Kentucky}, \orgaddress{\street{505 Rose Street}, \city{Lexington}, \postcode{40506}, \state{KY}, \country{USA}}}

\affil[3]{\orgdiv{SOFIA Science Center}, \orgname{NASA Ames Research Center}, \orgaddress{\street{M.S. N232-12 Moffett Field}, \city{Mountain View}, \postcode{94035}, \state{CA}, \country{USA}}}

\affil[4]{\orgname{Space Telescope Science Institute}, \orgaddress{\street{3700 San Martin Drive}, \city{Baltimore}, \postcode{21218}, \state{MD}, \country{USA}}}

\affil[5]{\orgdiv{Center for Astrophysics and Space Sciences, Department of Physics}, \orgname{University of California, San Diego}, \orgaddress{\street{9500 Gilman Drive}, \city{La Jolla}, \postcode{92093}, \state{CA}, \country{USA}}}

\affil[6]{\orgdiv{Pete V. Domenici Science Operations Center}, \orgname{National Radio Astronomy Observatory}, \orgaddress{\street{P.O. Box O, 1003 Lopezville Road}, \city{Socorro}, \postcode{87801-0387}, \state{NM}, \country{USA}}}

\affil[7]{\orgname{The Observatories of the Carnegie Institution for Science}, \orgaddress{\street{813 Santa Barbara Street}, \city{Pasadena}, \postcode{91101}, \state{CA}, \country{USA}}}

\affil[8]{\orgdiv{Steward Observatory}, \orgname{University of Arizona}, \orgaddress{\street{933 N Cherry Avenue}, \city{Tucson}, \postcode{85721}, \state{AZ}, \country{USA}}}

\affil[9]{\orgdiv{Department of Physics and Astronomy and George P. and Cynthia Woods Mitchell Institute for Fundamental Physics and Astronomy}, \orgname{Texas A\&M University}, \orgaddress{\street{576 University Drive}, \city{College Station}, \postcode{77843-4242}, \state{TX}, \country{USA}}}

\affil[10]{\orgdiv{Department of Physics and Astronomy}, \orgname{University of California, Los Angeles}, \orgaddress{\street{430 Portola Plaza}, \city{Los Angeles}, \postcode{90095}, \state{CA}, \country{USA}}}

\maketitle

\textbf{In \cite{chen23}, we introduced a new method to directly measure temperature fluctuations and applied it to a nearby dwarf galaxy, Mrk~71, finding a temperature fluctuation parameter $t^2 = 0.008\pm 0.043$. This result is lower by $\sim 2\sigma$ than the value required to explain the abundance discrepancy (AD) in this object. In the Matters Arising article submitted by M\'endez-Delgado et al., the authors claim that using the same data presented in \cite{chen23} in a different way, it is possible to conclude that the measurements are consistent with a larger $t^2 \simeq 0.1$ inferred indirectly from recombination lines (RLs). 
However, this requires a higher density such that the infrared \Oiii\ 52~\um\ and \Oiii\ 88~\um\ lines -- which form the basis of the direct measurement method -- are mutually inconsistent. Moreover, to reach agreement between the direct $t^2$ measurement and the larger $t^2$ value inferred from RLs requires systematically varying four parameters by $\sim 1\sigma$ from their best-determined values, 
which collectively amount to a $\sim2\sigma$ difference, 
consistent with the significance ($\sim 2 \sigma$) originally reported in \cite{chen23}. 
Therefore, we conclude that the results of \cite{chen23} hold, and that the combined optical and infrared \Oiii\ data disfavour $t^2 \simeq 0.1$ at the $\approx2\sigma$ level in Mrk~71. Future work is nonetheless warranted to better understand the AD associated with both optical and infrared emission line analysis.}


\section{Absolute Flux Calibration}

In their Matter Arising article, the authors questioned the accuracy of the absolute flux calibration in the optical and IR spectra. However, all systematic uncertainties discussed in their article have been accounted for in \cite{chen23}. 

As noted in \cite{chen23}, the Keck/KCWI data were taken under clear conditions with seeing $\lesssim 1"$. The flux was calibrated on photometric standard stars (G 47-18 and GD190) observed during the same night and on the HST/WFC3-UVIS F438W image. The absolute flux calibration accuracy of WFC3-UVIS is $< 0.5\%$ \cite{calamida22}. The standard deviation of the spectral variation between the observed and calibration spectra of the standard stars is $<3\%$. We adopted the 3\% accuracy. 

For SOFIA/FIFI-LS, the absolute flux calibration accuracy is established to be $\lesssim 10\%$ \cite{fadda23}. The Matters Arising manuscript notes the presence of telluric features. However, there is only one small ($\gtrsim$80\% transmission) feature lying 137 km/s away from the \Oiii~52~\um\ line (which has an intrisic width of $\sigma \simeq 42$ \kms). The telluric feature has a negligible effect ($<1\%$) on the measured flux.  
For Herschel/PACS, the absolute flux calibration accuracy of the B2B band is $\sim$12\% (\url{http://herschel.esac.esa.int/Docs/PACS/html/ch04s10.html}). These absolute flux calibration uncertainties for SOFIA and Herschel are consistent with the 15\% relative flux uncertainty \cite{fadda16, sutter22} established by directly comparing measurements of the same features. In \cite{chen23}, we quadratically divided the 15\% relative flux uncertainty resulting in 11\% uncertainty for both SOFIA/FIFI-LS and Herschel/PACS, similar to their individual estimates above. The results are essentially unchanged if we adopt 10--12\% or otherwise separate the SOFIA and Herschel flux uncertainties.

\section{Electron Density}

The electron density ($n_e$) has a significant effect on the $t^2$ measurement because the \Oiii\ IR line emissivity is $n_e$-dependent. 
Our analysis adopts the value $n_e = 140_{-70}^{+50}~\mathrm{cm}^{-3}$ measured from the \Oiii~52~\um/88~\um\ ratio, which at face value is the most appropriate for this purpose. The uncertainty in $n_e$ accounts for the 15\% systematic uncertainty associated with the flux calibration discussed above and in the Matters Arising article. 
There appears to be a misunderstanding in the Matters Arising manuscript which states that we adopted a value close to $n_e = 160\pm 10~\mathrm{cm}^{-3}$ measured from the \Oii~3726/3729 ratio. In fact, the density independently derived from \Oii\ is not used at all in the measurement of O$^{++}$/H$^+$ abundance or $t^2$. Our conclusions in \cite{chen23} are based \textit{entirely} on the \Oiii\ emission lines. 

We emphasize that a main motivation of our new method presented in \cite{chen23} is to \textit{directly} and \textit{self-consistently} measure $t^2$ using only the collisionally excited lines (CELs) of O$^{++}$. It is thus intentionally agnostic to the densities derived from \Oii, \Oiirec~RLs, and other diagnostics. We show these in Figure~3 of \cite{chen23} mainly for purposes of comparison and reference. Adopting $n_e$(\Oiirec~RLs) as suggested by M\'endez-Delgado et al. is against the physical motivation of our study. 

As discussed in \cite{chen23}, we recognise that high-density gas could potentially affect the metallicity measured from \Oiii\ IR lines. However, it requires a significant presense ($> 10\%$) of gas with a density higher than the critical density of the \Oiii\ IR emission to significantly modulate the \Oiii\ IR temperature and metallicity. We do not see evidence supporting the significant presence of such gas from various density indicators in Mrk~71. 
Further quantifying the magnitude of density fluctations is beyond the scope of the present work and deserves greater attention in the field.

\section{Temperature Fluctuations}

In the Matter Arising manuscript and their Fig. 1, the authors concluded that the $t^2$ value derived from \Oiii~CELs can be consistent with the value $t^2 \simeq 0.097$ required to reconcile the AD, using the same data presented in \cite{chen23}. However, we find that their analysis is not contradictory to our conclusion in \cite{chen23}, and in fact reinforces the fact that the direct measurement of $t^2$ is different by $\sim2\sigma$ from that inferred from the AD. 

In their independent analysis, M\'endez-Delgado et al. used the combined \Oiii\ 52~\um + 88~\um\ flux. This approach discards the joint constraining power to treat the \Oiii\ 52~\um\ and 88~\um\ independently, and does not seem well-justified given the reasonably good characterization of measurement uncertainties (including absolute flux calibration) and especially the fact that the flux \textit{ratio} \Oiii\ 52~\um/88~\um\ is sensitive to density as discussed above. Using the summed 52~\um + 88~\um\ flux with a higher density of $n_e = 310~\mathrm{cm}^{-3}$, as argued for in the Matters Arising article, in fact requires tension with the observations in that the individual infrared \Oiii\ line fluxes are not in mutual agreement with such a high density.

In order to explain a value $t^2 \simeq 0.097$ predicted from the AD, several measurements from \cite{chen23} must be systematically varied by $\gtrsim 1\sigma$ from their best determined values. M\'endez-Delgado et al. show that this can be achieved by taking $\chb = 0.05$ ($1\sigma$ from our best determined \chb), combined with $n_e = 310 \pm 50~\mathrm{cm}^{-3}$ ($> 2\sigma$ from our best determined $n_e$; alternatively this can be thought of as $\gtrsim \sqrt{2} \simeq 1.4 \sigma$ differences in both the \Oiii~52~\um\ and 88~\um\ line fluxes), to reach a marginal $\sim 1\sigma$ consistency between $T_e$(measured from IR and CEL) and $T_e$(predicted from optical RL and CEL). 
This scenario can be equivalently expressed as a $\sim2\sigma$ difference between $t^2$ directly measured from the far-IR lines compared to the $t^2$ value predicted from the AD.
A difference of $\sim1\sigma$ \textit{in four independent parameters} relative to their best-determined values collectively represents a $\sim 2\sigma$ inconsistency between the two $t^2$ values, i.e., a difference of $1\sigma$ in each of $N$ independent parameters amounts to a total difference of $\sqrt{N} \sigma$ from standard error propagation. In this case \chb, \Oiii~52~\um\ flux, \Oiii~88~\um\ flux, and $T_e$(predicted) represent $N=4$ independent parameters.
This level of tension is in excellent agreement with the $2.1\sigma$ difference that we report between directly measured versus AD-predicted $t^2$ values in \cite{chen23}. We conclude that the analysis of M\'endez-Delgado et al. reinforces our conclusion of a $\sim2\sigma$ discrepancy. 

We acknowledge that in \cite{chen23}, the direct $t^2$ measurement was only conducted on one object, Mrk~71. To what extent temperature fluctuations contribute to the measured AD in the general population of \Hii\ regions requires further research with larger samples. 
We also acknowledge that fluctuations of other physical properties (e.g., $n_e$, \chb) can potentially affect the $t^2$ measurements. Fluctuations in $n_e$ can lead to an underestimate of $t^2$ from our infrared-line methodology, while fluctuations in dust attenuation can have the opposite effect. 
Our group is actively researching these and related topics. Meanwhile, we thank the authors of the Matter Arising submission and the broader scientific community for their interest in our work. 
We believe that the method of directly measuring $t^2$ from the collisionally excited lines holds great promise for understanding both the cause of the AD and the temperature structure of ionised nebulae. 

 {\textit{Correspondence and requests for materials should be addressed to Y. Chen.}}

\noindent\textbf{Author Contributions}

Y.C. led the preparation of the manuscript. All authors provided feedback to the manuscript. 

\noindent\textbf{Data availability} 

All data have been provided in \cite{chen23}.

\noindent\textbf{Code availability}

This study uses publicly available software/packages that have been provided in \cite{chen23}.

\noindent\textbf{Competing Interest Statement}
The authors declare no competing interest.


\end{document}